\documentclass{kluwer}    
\usepackage[dvips]{graphicx}

\newdisplay{guess}{Conjecture}
\newcommand{\etal}{{\it{et~al.~}}}
\def\yrmu{\hbox{\,yr$^{-1}$}~}
\def\mum{\hbox{\,$\mu$m\,}}
\def\sol{\hbox{$_{\odot}$~}}
\def\ir{\hbox{$_{\rm IR}$~}}

\def\sm{\hbox{$\sim$}}
\begin{document}
\begin{article}
\begin{opening}         
\title{Understanding galaxy formation with ISO deep surveys}
\author{David \surname{Elbaz}}
\runningauthor{David Elbaz}
\runningtitle{Understanding galaxy formation with ISO deep surveys}
\institute{CEA Saclay$/$DSM$/$DAPNIA$/$Service d'Astrophysique\\
Orme des Merisiers, F-91191 Gif-sur-Yvette Cedex\\
France}
\date{July 26, 2004}

\begin{abstract}
We present the results obtained through the various ISO extragalactic
deep surveys. While IRAS revealed the existence of galaxies forming
stars at a rate of a few tens (LIRGs) or even hundreds (ULIRGs) solar
masses in the local universe, ISO not only discovered that these
galaxies were already in place at redshift one, but also that they are
not the extreme objects that we once believed them to be. Instead they
appear to play a dominant role in shaping present-day galaxies as
reflected by their role in the cosmic history of star formation and in
producing the cosmic infrared background detected by the COBE
satellite in the far infrared to sub-millimeter range.
\end{abstract}
\end{opening}           

\section{General perspective}
\label{general}
It is widely accepted that in the local universe stars form in giant
molecular clouds (GMCs) where their optical and mostly UV light is
strongly absorbed by the dust which surrounds them. Whether extinction
was already taking place in the more distant universe where galaxies
are less metal rich was less obvious ten years ago. Galaxies forming
stars at a rate larger than about 20 M\sol\yrmu were known to radiate
the bulk of their luminosity above 5\mum thanks to IRAS, the so-called
luminous (LIRGs, 12 $>~log($L\ir$/L\sol)\geq$ 11) and ultra-luminous
(ULIRGs, $log($L\ir$/L\sol)\geq$ 12) infrared (IR) galaxies. In the
past, when galaxies were more gaseous and formed the bulk of their
present-day stars, it would have been logical to expect to detect the
past star formation events of galaxies in the IR regime and to detect
a large population of LIRGs/ULIRGs too. However, prior to the launch
of the Infrared Space Observatory (ISO, Kessler \etal 1996), this idea
was not widely spread. Partly because of a cultural reason: star
formation rates (SFR) were commonly measured from optical emission
lines and rest-frame UV light in galaxies. This may explain why the
redshift evolution of the SFR density per unit comoving volume
computed by Madau \etal (1996) became famous. However in the first
presentation of the density of UV light per unit comoving volume
(Lilly \etal 1996) the authors were cautious to avoid converting their
UV light into a SFR because of the unknown factor to correct for
extinction. Already IRAS observations indicated a rapid decline of the
comoving number density of ULIRGs since $z$\sm0.3 (Kim \& Sanders
1998, see also Oliver \etal 1996), but this was over a small redshift
range and with small number statistics.  In the few years that
followed the launch of ISO, several observations showed that galaxy
formation could not be understood, at least on an observational basis,
without accounting for dust extinction as a major ingredient. The ISO
deep surveys played a major role in this process, together with other
results summarized below. They clearly established that extreme events
such as those taking place in local LIRGs and ULIRGs must have been
more common in the past, so much that they can now be considered as
a standard phase that most galaxies experienced during their lifetime,
at least once, but maybe even several times.

The first result of the ISOCAM surveys, as well as the ISOPHOT ones,
was the great difference of the counts measured at faint flux
densities with respect to local ones from IRAS (Elbaz \etal 1999, Dole
\etal 2001). The universe must have been much richer in IR luminous
galaxies in the past, either because galaxies were more IR luminous,
at fixed galaxy density, and/or because the number density of galaxies
was larger in the past, which was partly expected due to the reverse
effect of hierarchical galaxy formation through mergers. The strength
of the excess of faint objects came as a surprise, but its
consequences on the past star formation history of galaxies was
confirmed by the convergence of other observations going in the same
direction:

- the nearly simultaneous discovery of the cosmic infrared background
(CIRB, Puget \etal 1996, Fixsen \etal 1998, Hauser \& Dwek 2001 and
references therein), at least as strong as the UV-optical-near IR one,
whereas local galaxies only radiate about 30\,\% of their bolometric
luminosity in the IR above $\lambda\sim$ 5\mum.

- the 850\mum number counts from the SCUBA sub-millimeter bolometer
array at the JCMT (Hughes \etal 1998, Barger \etal 1998, Smail \etal
2002, Chapman \etal 2003, and references therein) which also indicate
a strong excess of faint objects in this wavelength range, implying
that even at large redshifts dust emission must have been very large
in at least the most active galaxies.

- the most distant galaxies, individually detected thanks to the
photometric redshift technique using their Balmer or Lyman break
signature showed the signature of a strong dust extinction. The
so-called ``$\beta$-slope'' technique (Meurer \etal 1999) used to
derive the intrinsic luminosity of these galaxies and correct their UV
luminosity by factors of a few (typically between 3 and 7, Steidel
\etal 1999, Adelberger \& Steidel 2000) was later on shown to even
underestimate the SFR of LIRGs/ULIRGs (Goldader \etal 2002).

- the slope of the sub-mJy deep radio surveys (Haarsma \etal 2000).

It has now become clear that the cosmic history of star formation
based on rest-frame UV or emission line indicators of star formation
such as [0II] or [H$\alpha$] strongly underestimate the true activity
of galaxies in the past if not corrected by strong factors due to dust
extinction. Although distant galaxies were less metal rich and much
younger, they must have found the time to produce dust rapidly in
order to efficiently absorb the UV light of their young
stars.

We have tried to summarize in the following the role played by the ISO
deep surveys in establishing this new perspective (see also Genzel \&
Cesarsky 2000). However, ten years after ISO's launch we are still
trying to understand the consequences of these findings on galaxy
formation scenarios. Are these distant LIRGs really similar to local
ones ? What do they teach us about the connection between star and
galaxy formation on one side (IMF, triggering of star formation,
conditions of star formation,...) and between galaxy and large-scale
structure formation on the other side (role of the environment in
triggering star formation events, galaxy versus group and cluster
formation, ellipticals versus spirals, ...) ?  How much energy
radiated by compton thick embedded active nuclei remains undetected
even by the Chandra and XMM-Newton X-ray observatories ? These
questions together with others that will be discussed in the following
demonstrate the liveliness of this field that will continue to feed
the next generation of telescopes and instruments to come such as
Herschel, ALMA, the James Webb Space Telescope (JWST) or the Spitzer
and GALEX space observatories presently in use.

\begin{figure} 
\includegraphics[width=8cm]{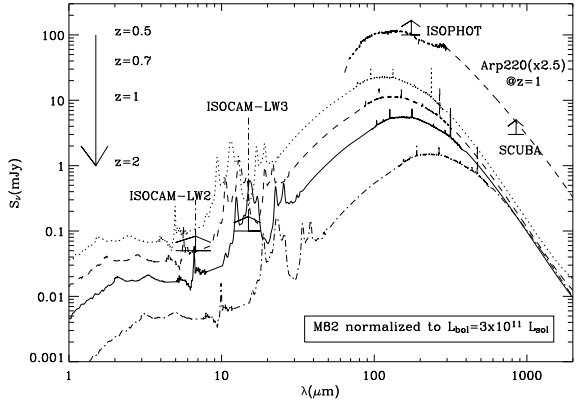}
\caption{Effect of the k-correction on the detection of distant LIRGs
and ULIRGs by ISOCAM, ISOPHOT and SCUBA. Only ISOCAM at 15\mum can
detect LIRGs up to $z\sim$ 1, while ISOPHOT and SCUBA are sensitive to
distant ULIRGs of a few 10$^{12}$ L\sol.}
\label{FIG:seds}
\end{figure}

\section{The ISO surveys}
\begin{figure}
\includegraphics[width=8cm]{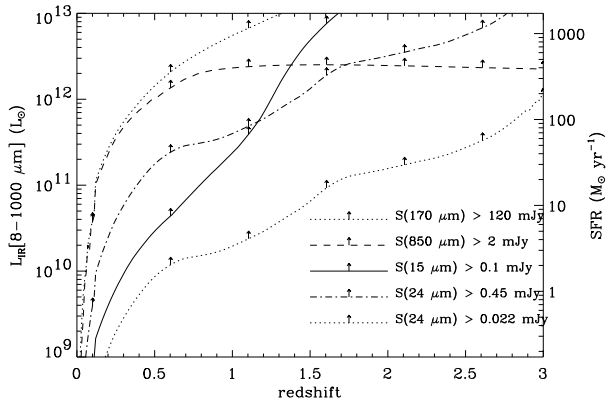}
\caption{Sensitivity limits of ISOCAM (15\mum, 0.1 mJy), ISOPHOT
(170\mum, 120 mJy), SCUBA (850\mum, 2mJy) and Spitzer (24\mum, 0.45
mJy and 0.022 mJy corresponding to the expected detection limits of
the SWIRE and GOODS Legacy Programs). This figure was generated
assuming that distant galaxies SEDs are similar to local ones. We used
the library of template SEDs constructed by Chary \& Elbaz (2001).}
\label{FIG:sens}
\end{figure}
Extragalactic deep surveys with ISO were performed in the mid and far
IR with ISOCAM (Cesarsky \etal 1996) and ISOPHOT (Lemke \etal 1996)
respectively. In both wavelength ranges, the steep slopes of the
number counts indicate that a rapid decline of the IR emission of
galaxies must have taken place from around $z\sim$1 to $z=$0. As shown
in Fig.~\ref{FIG:seds}, ISOCAM could detect galaxies at 15\mum in the
LIRG regime up to $z\sim$1.3, while ISOPHOT was limited to either
nearby galaxies or moderately distant ULIRGs such as the one of L\ir
$\sim~4\times10^{12}$ L\sol at a redshift of $z$=1 shown in the
Fig.~\ref{FIG:seds} (normalized SED of Arp 220). The detection limits
of ISOCAM and ISOPHOT are compared to SCUBA and Spitzer in the
Fig.~\ref{FIG:sens}.  The deepest ISO surveys reached a completeness
limit of 0.1 mJy at 15\mum (plain line) and a depth of $S_{15}\sim$
40\,$\mu$Jy (incomplete, Aussel \etal 1999) in blank fields or a twice
deeper completeness level in the central part of nearby galaxy
clusters using lensing magnification (Altieri \etal 1999, Metcalfe
\etal 2003), and 120 mJy at 170\mum (dotted line). The right axis on
the plot shows the conversion of the IR detection limit into a SFR
detection limit. Any galaxy forming stars at a rate larger than 30
M\sol\yrmu could be detected with ISOCAM up to $z\sim$ 1, assuming
that the measured 15\mum flux density of a galaxy can be used to
derive its ``bolometric'' IR luminosity, i.e. L\ir=L(8-1000\mum) (see
Sect.~\ref{MIRtracer}).

Number counts represent the first scientific result of an extragalactic
survey. They can be used to constrain the models. Used alone, they
leave some degeneracies unsolved but at least they can demonstrate
whether the distant universe was different from the local one in this
wavelength range by comparing them to ``no evolution'' predictions
assuming some spectral energy distributions (SEDs) for the
k-correction. 

\section{The ISOCAM 6.75\mum deep surveys}
Deep images of blank fields at 6.75 and 15\mum done with
ISOCAM provide a different view on the distant universe. While the
ISOCAM-LW3 filter, centered at 15\mum, can probe dust emission up to
$z\sim$ 2 for luminous objects, the redshift range to probe star
formation from dust emission with the ISOCAM-LW2 band, centered at
6.75\mum, is limited to the relatively nearby universe due to
k-correction. However the emission of the old stellar component, which
peaks in the near IR, being brought to this wavelength range for
high-$z$ galaxies, their stellar masses are better constrained from
this flux density. 

Sato \etal (2004) derived stellar masses around $M_{\star}\sim$
10$^{11}$ M$_{\odot}$ for galaxies with $z\sim$ 0.2-3 from the
correlation between rest-frame near IR luminosity and stellar mass.
The stellar mass-to-light ratios were derived from model fit of the
set of observed magnitudes depending on the galaxies star formation
histories. 

\begin{figure}
\includegraphics[width=12cm]{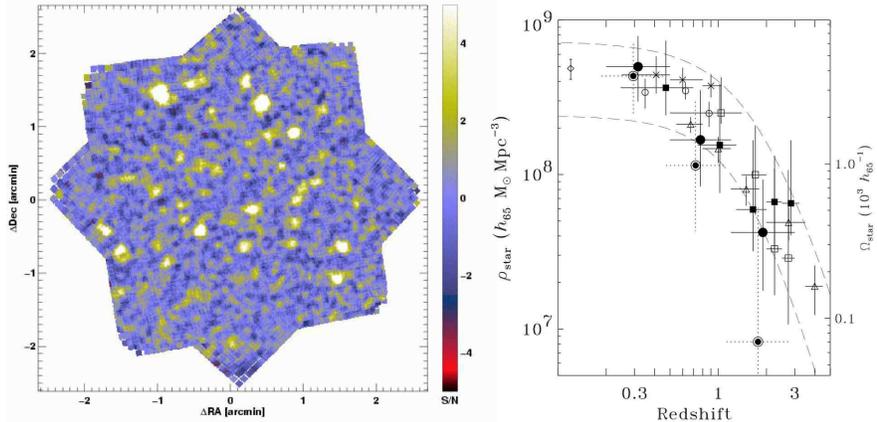}
\caption{{\bf (left)} ISOCAM 6.75\mum image of the SSA13 field (Fig.2
from Sato \etal 2003, 23 hour observation). This image reaches an
80\,\% completeness limit of 16\,$\mu$Jy in the central 7 arcmin$^2$
region. North is top and east to the left in J2000.  The center is
(RA, Dec) $=$ ($13^{\rm h}\,12^{\rm m}\,26^{\rm s},
42\,^{\rm o}44'\,24''.8$).  The map shows signal-to-noise ratio
(S/N) per 0.6 arcsec sub-pixel.  The signal is an average, weighted by
the inverse of the assigned variance, and the noise is a normalized
standard deviation. {\bf (right)} Stellar mass density in the universe
as a function of redshift (Fig.13 from Sato \etal 2004).  The right
axis shows densities normalized to the critical density of the
universe.  The contributions of the 6.75\,$\mu$m galaxies are shown
with solid and double circles for the ``combined'' (including
photometric redshifts) and purely spectroscopic samples, respectively.
The double circles are plotted at slightly lower redshifts.  The
horizontal bars represent the redshift ranges of the bins and the
vertical bars show one sigma errors, taking account of Poisson noise
and uncertainties in stellar mass and $V_{\rm max}$.  Several other
estimates are overlaid (see Sato \etal 2004 for references).  The
empty/solid squares and the diamond are obtained from full integration
of a Schechter fit to their respective luminosity or stellar mass
function at each redshift bin.  The X marks and the empty circles are
quasi-fully integrated values with a finite integration range from
10\,$L^*$ to 1/20\,$L^*$, and $10.5<\log (M_{\rm star}
[h_{65}^{-2}$\,M$_\odot])<11.6$, respectively.  The triangles are
simply summed values of the detected sources.  The two dashed curves
are deduced by integrating the star formation rate density in the
universe, which is derived from the UV luminosity density as a
function of redshift (Cole \etal 2001).  The upper curve is an
extinction corrected case for $E(B-V)=0.15$, and the lower one has no
dust correction.}
\label{FIG:sato}
\end{figure}

The contribution of the 6.75\mum selected galaxies to the cosmic
stellar mass density of galaxies per unit comoving volume as a
function of redshift were estimated to be comparable to those inferred
from observations of UV bright galaxies (see
Fig.~\ref{FIG:sato}). Given the narrow mass ranges, these estimates
were obtained as simple summations of the detected sources and should
therefore be considered as lower limits only. The faint 6.7\,$\mu$m
galaxies generally had red colors.  A comparison with a particular
population synthesis model suggests that they have experienced
vigorous star formation at high redshifts.  The derived large stellar
masses for the faint 6.7\,$\mu$m galaxies also support such star
forming events at the past. However these events can only be detected
at longer wavelengths using the 15\mum ISOCAM or 90 and 170\mum
ISOPHOT bands with ISO (or SCUBA at 850\mum and now the Spitzer Space
Telescope). The next sections are devoted to these surveys and their
consequences on our understanding of the history of star formation
in the universe.

\section{The ISOPHOT far IR deep surveys}
\begin{figure*} 
\centering
\includegraphics[width=12cm]{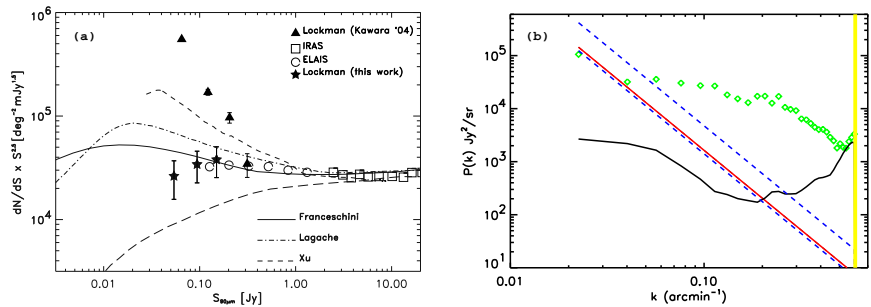}
\caption{{\bf (a)} Differential 90 {\rm $\mu$m} counts dN/dS
normalized to the Euclidean law ($N \propto S^{-2.5}$) (extracted from
Rodighiero \& Franceschini 2004). Results compared with those from the
preliminary analysis of the ISOPHOT ELAIS survey (Efsthatiou \etal
2000, open circles) and with those from Kawara \etal (2004, filled
triangles).  The long-dashed line shows the expected contribution of
non-evolving spirals as in the model of Franceschini \etal
(2001). {\bf (a)} Fluctuations of the CIB in a power spectrum analysis
of the FIRBACK/ELAIS N2 field at 170 microns by Puget \& Lagache
(2000). Observed power spectrum: diamond; straight continuous line:
the best fit cirrus power spectrum; dash line: cirrus power spectrum
deduced from Miville-Desch\^enes \etal (2002); continuous curve:
detector noise.}
\label{FIG:rodh} 
\end{figure*} 
ISOPHOT was used to survey the sky at 90 and 170\mum, mainly (plus a
few surveys at 60, 120, 150 and 180 $\mu m$, see Juvela \etal 2000;
Linden-Voernle \etal 2000; cf review by Dole 2002). At 90\mum,
the 46 arcseconds pixels and FWHM of ISOPHOT represented a significant
improvement with respect to IRAS, however due to k-correction and
sensitivity limitations, only the local universe could be probed at
such wavelengths. The 170\mum band was more favorable for the
detection of distant ULIRGs as shown in the Fig.~\ref{FIG:seds} where
the peak emission of a ULIRG of $4\times10^{12}$ L\sol and redshifted
at z=1 is shown (normalized SED of Arp 220). A major issue at these
wavelengths with a 60 cm telescope is obviously the identification of
optical counterparts for the determination of a redshift and the
separation of local and distant galaxies.

\subsection{Source Counts and Cosmic IR Background}
Source counts at 170 $\mu m$ (e.g. Kawara \etal, 1998; Puget \etal,
1999; Dole \etal, 2001) exhibit a steep slope of $\alpha = 3.3 \pm
0.6$ between 180 and 500 mJy and, like in the mid IR range, show
sources in excess by a factor of 10 compared with no evolution
scenario. The brightness fluctuations in the Lockman Hole were used by
Matsuhara et al (2000) to constrain galaxy number counts down to 35
mJy at 90\mum and 60 mJy at 170\mum, confirming the existence of a
strong evolution down to these flux densities. Using a new data
reduction method, Rodighiero \& Franceschini (2004) extended the
previous works of Kawara \etal (1998), Efstathiou \etal (2000) and
Linden-Voernle \etal (2000) down to lower flux densities (30 mJy at
90\mum) and found a clear excess of faint objects with respect to no
evolution (see Fig.~\ref{FIG:rodh}a).  However, the resolved sources
account for less than 10\,\% of the Cosmic Infrared Background at
170\mum, which is expected to be resolved into sources in the 1 to 10
mJy range.

Sources below the detection limit of a survey create fluctuations. If
the detection limit does not allow to resolve the sources dominating
the CIB intensity, which is the case in the far IR with ISO,
characterizing these fluctuations can constrain the spatial
correlations of these unresolved sources of cosmological
significance. An example of the modeled redshift distribution of the
unresolved sources at 170 $\mu m$ can be found in Fig.~12 of Lagache
\etal (2003); the sources dominating the CIB fluctuations have a
redshift distribution peaking at $z \sim 0.9$.  After the pioneering
work of Herbstmeierer \etal (1998) with ISOPHOT, Lagache \& Puget
(2000) discovered them at 170 $\mu m$ in the FIRBACK data, followed by
other works at 170 and 90 $\mu m$ (Matsuhara \etal 2000; Puget \&
Lagache, 2000; Kiss \etal 2001). Fig.~\ref{FIG:rodh}b shows the CIB
fluctuations in the FN2 field by Puget \& Lagache, (2000), at
wavenumbers $0.07 < k < 0.4$ arcmin$^{-1}$.

\subsection{Nature of the ISOPHOT Galaxies}
Determining the nature of the far IR galaxies has been a longer
process than in the mid IR, mainly because of the difficulty to find
the shorter wavelength counterparts in a large beam. Various
techniques have been used to overcome this problem, one of the most
successful being the identification using 20 cm radio data
(e.g. Ciliegi \etal 1999). Another technique is the far IR
multiwavelength approach (Juvela \etal 2000) that helps constraining
the position and the SED; it also helps to separate the cirrus
structures from the extragalactic sources. A variation is to use
ISOCAM and ISOPHOT data, like the ELAIS Survey (Rowan-Robinson \etal
2004, see also Oliver in this book).  Finally, the Serendipity Survey
(Stickel \etal 1998, 2000), by covering large and shallow areas,
allows to detect many bright objects easier to follow-up or already
known.

Far IR ISO galaxies can be sorted schematically into two populations.
First, the low redshift sources, typically $z<0.3$ (e.g Serjeant \etal
2001; Patris \etal 2002, Kakazu \etal 2002), have moderate IR
luminosities, below $10^{11} L_{\odot}$, and are cold (Stickel \etal
2000). Second, sources at higher redshift, $z \sim 0.3$ (Patris et
al., 2002) and beyond, $z \sim 0.9$ (Chapman \etal 2002) are more
luminous, typically $L > 10^{11} L_{\odot}$, and appear to be
cold. Serjeant \etal (2001) derived the Luminosity Function at 90 $\mu
m$, and started to detect an evolution compared to the local IRAS 100
$\mu m$ sample.

\section{The ISOCAM 15\mum deep surveys}
\begin{figure*} 
\centering
\includegraphics[width=12cm]{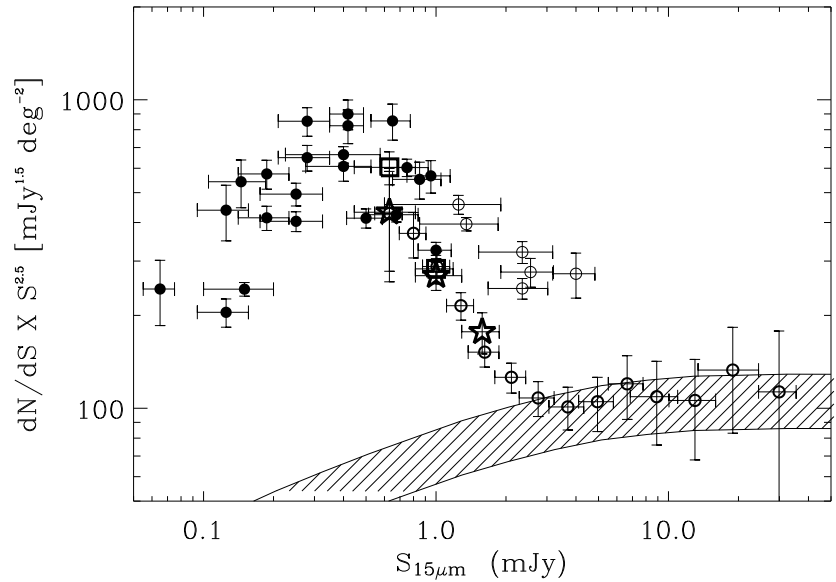}
\caption{ISOCAM 15\mum differential counts, with 68\% error bars. The
counts are normalized to a Euclidean distribution of non-evolving
sources, which would have a slope of $\alpha=-2.5$ in such a
universe. This plot is an extension of the Fig.2 from Elbaz \etal
(1999) combining the data points from Elbaz \etal (1999), for the
ISOCAM Garanteed Time Extragalactic Surveys (IGTES, filled and empty
dots for sources below and above 1 mJy respectively, see text),
Gruppioni \etal (2003, bold empty circles) for the ELAIS-S1 field,
Rodighiero \etal (2004, large open stars) and Fadda \etal (2004, large
open squares), for the Shallow and Deep Surveys of the Lockman Hole
respectively (analysis of the IGTES data using the ``Lari'' technique,
see Sect.~\ref{isocam}). The hatched area materializes the range of
possible expectations from models assuming no evolution and normalized
to the 15\,$\mu$m local luminosity function (LLF) from Fang \etal
(1998) and the template SED of M51 (with an assumed uncertainty of
$\pm$ 20\,\%; this 15\,$\mu$m LLF was recently confirmed over a larger
sample of local galaxies by H.Aussel, private communication).}
\label{FIG:dNdS} 
\end{figure*} 

\label{isocam}
{\scriptsize 
\begin{table*}
\begin{center}
\leavevmode
\vspace{0.5 em}
\caption{\em Table of the ISOCAM extragalactic surveys sorted by
increasing depth. Col.(1) Survey name. Col. (2) wavelength of the
survey with ``7'' for the LW2 filter centered at 6.75\,$\mu$m and
covering 5-8.5\,$\mu$m, and ``15'' for the LW3 filter centered
15\,$\mu$m and covering 12-18\,$\mu$m. Col.(3) total area in square
arcmin. Col.(4) integration time per sky position. Col.(5) depth (and
80\,$\%$ completeness limit when indicated). Col.(6) number of objects
detected above this depth.}
\begin{tabular}{|l|ccccc|}
\hline
Name              &$\lambda~(\mu m)$& Area ('$^2$)& Int.(min)& depth(mJy)        &\#~objects \\
\hline										     
ELAIS N1$^1$      &   15	    &9612         & 0.7;0.7  & 1;0.7	         & 490       \\
ELAIS N2$^1$      & 7;15	    &9612;9612    & 0.7;0.7  & 1;0.7	         & 628;566   \\
ELAIS N3$^1$      & 7;15	    &4752;3168    & 0.7;0.7  & 1;0.7	         & 189;131   \\
ELAIS S1$^1$      & 7;15	    &6336;14256   & 0.7;0.7  & 1;0.7	         & 304;317   \\
ELAIS S2$^1$      & 7;15	    &432;432      & 0.7;0.7  & 1;0.7	         & 40;43     \\
Lockman Shallow$^{2,3}$& 15         & 1944        &   3      &       0.25        &       457 \\ 
                  &      	    &(80\,\% compl.)&        &       0.45        &       260 \\
Comet Field$^4$   & 12              & 360         &   10     &       0.5         &        37 \\
CFRS14+52$^5$     & 7;15	    & 100;100     & 18;11    & 0.3  ;0.4         &   23;41   \\
CFRS03+00$^6$     & 7;15	    & 100;100     &  6;22    & 0.5  ;0.3         &   -       \\
Lockman Deep$^{2,7}$& 7;15          & 500;500     & 18;11    & 0.3  ;0.4         &   166     \\
Marano DSF$^2$    & 7;15            & 900;900     & 15.4;15.4& 0.19 ;0.32        &   180     \\
A370$^{8}$        & 7;15	    & 31.3;31.3   & 42;42    &0.052;0.21$^{(u)}$ &    4;20   \\
                  &      	    &(80\,\% compl.)&        &0.080;0.293$^{(u)}$&           \\
Marano UDSR$^2$   & 7;15            & 85;90       &120;114   & 0.18 ;0.14        &    -;142  \\
Marano UDSF$^2$   & 7;15            & 89;90       &114       & 0.08 ;0.14        &  115;137  \\
A2218$^8$         & 7;15            & 20.5;20.5   &84;84     &0.054;0.121$^{(u)}$&   18;46   \\
                  &      	    &(80\,\% compl.)&        &0.079;0.167$^{(u)}$&           \\
HDFN$+$FF$^{2,9}$ & 7;15	    & 10;27	  & 116;135  & 0.05 ;0.1         &    7;44   \\
HDFS$^{2,10}$     & 7;15	    & 28;28	  & 168;168  & 0.05 ;0.1         &   16;63   \\
A2390$^{11,8}$    & 7;15	    & 5.3;5.3     & 432;432  &0.038;0.050$^{(u)}$&   10;28   \\
                  &      	    &(80\,\% compl.)&          &0.052;0.092$^{(u)}$&         \\
Lockman PGPQ$^{12}$& 7	            & 9           & 744      &       0.034       &   15      \\
SSA13$^{13}$      & 7	            & 16          & 1264     &       0.006       &      65   \\
                  & 	            &7(80\,\% compl.)&       &       0.016       &           \\ 
\hline        
\end{tabular}
\label{TAB:surveys}
\end{center}
\noindent{\footnotesize {\em References}: (1) Oliver \etal 2000,
Rowan-Robinson \etal 2004, (2) Elbaz \etal 1999 and in prep., (3)
Rodighiero \etal 2004, (4) Clements et al. 1999, (5) Flores \etal
1999, (6) Flores (private communication), (7) Fadda \etal 2004, (8)
Metcalfe \etal 2003, (9) Aussel \etal 1999, Goldschmidt \etal 1997,
(10) Oliver \etal 2002, (11) Altieri \etal 1999, (12) Taniguchi et
al. 1997, (13) Sato \etal 2003, Sato \etal 2004. The depth of the deep
surveys in the field of lensing clusters does not include the
correction for lensing amplification, as indicated by the $^{(u)}$}
\end{table*}
}
\begin{figure}
\includegraphics[width=8cm]{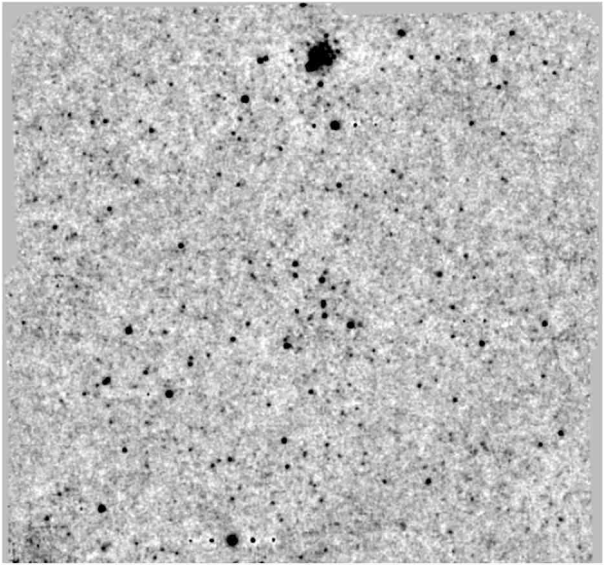}
\caption{ISOCAM 15\mum image of the Marano FIRBACK Deep Survey (DSF)
from the IGTES (Elbaz \etal 1999, and in prep.). Data reduction with
PRETI.}
\label{FIG:dsf}
\end{figure}

\begin{figure}
\includegraphics[width=8cm]{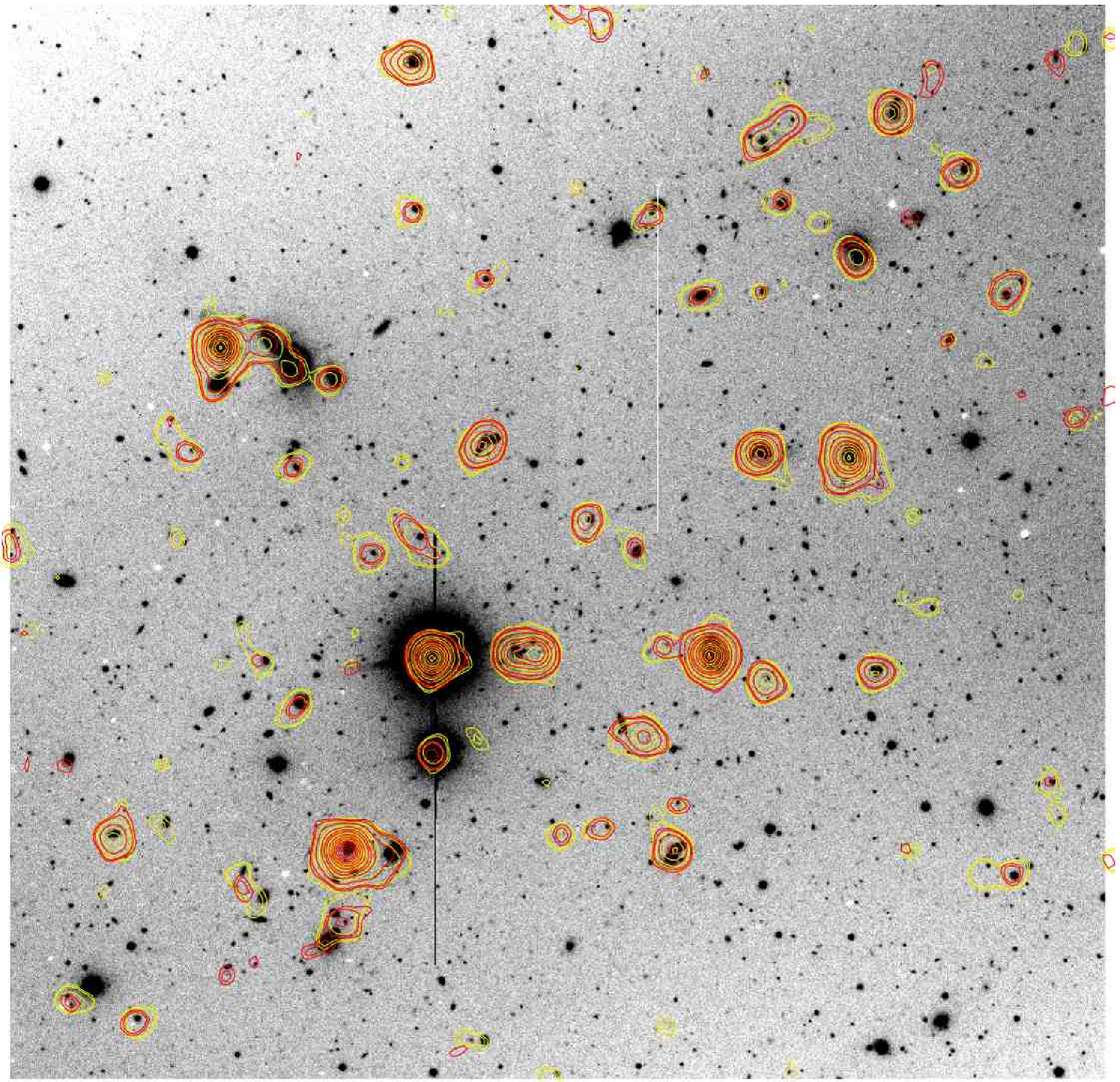}
\caption{ISOCAM 15\mum contours overlayed on the VLT-FORS2 image
(7'$\times$7', R-band) of the Marano FIRBACK Ultra Deep Survey (UDSF).
The LARI (light contours, yellow on screen, grey on paper) and PRETI
(dark contours, red on screen) detect the same objects.}
\label{FIG:laripreti}
\end{figure}

A series of surveys were performed within the Garanteed Time (IGTES,
ISOCAM Garanteed Time Extragalactic Surveys, Elbaz \etal 1999) and
Open Time as summarized in the Table~\ref{TAB:surveys}, where the
surveys at 7 (6.75\mum) and 15\mum are sorted by increasing depth
irrespective of wavelength. The major strength of ISOCAM is its
spatial resolution (PSF FWHM of 4.5'' at 15\mum, Okumura 1998) and
sensitivity, which permitted to detect galaxies down to the LIRG
regime up to $z\sim$1.3 in the deepest surveys
(Figs.~\ref{FIG:seds},~\ref{FIG:sens}), well above confusion.  Cosmic
rays were a stronger limitation than photon or readout noise by
inducing ghost sources when they were not perfectly removed,
especially those with long term transients associated to them.  Two
techniques were developped in order to solve this issue, the so-called
PRETI (Pattern REcognition Technique for Isocam data, Starck \etal
1999) and LARI (Lari \etal 2001) techniques. A third technique was
developped by D\'esert \etal (1999), in which the mosaic data were
analyzed using the beam switching approach, the Three Beam
technique. PRETI consists in a multi-scale wavelet decomposition of
the signal, while LARI tries to account for the physical processes
taking place in the detectors, including the effect of neighboring
pixels. The LARI technique was first applied to the ELAIS surveys (see
Oliver \etal, in these book) and more recently to the IGTES surveys of
the Lockman Hole (Rodighiero \etal 2004, Fadda \etal 2004) and of the
Marano field (Elbaz \etal, in prep.).  The quality of the ISOCAM
images is shown with the case of the Marano FIRBACK deep field
(30'$\times$30') in the Fig.~\ref{FIG:dsf}. The central part of this
field was imaged at a deeper level (UDSF) and analyzed with both
techniques. The resulting contours are overlayed on a VLT-FORS2 image
of the field, demonstrating the consistency and robustness of both
source detection algorithms (Fig.~\ref{FIG:laripreti}).

Lensing appeared to be a very powerful tool to extend the detection of
faint objects by a factor 2-3 (see Metcalfe \etal 2003). However
studies such as the relative clustering of IR galaxies versus field
galaxies are better done in larger homogeneous fields, where the role
of cosmic variance can also be quantified.  As we will see below, the
environment of distant LIRGs seems to play a major role in triggering
their star formation activity.

\subsection{Source Counts and Cosmic IR Background}
\label{camcounts}

\begin{figure}
\includegraphics[width=12cm]{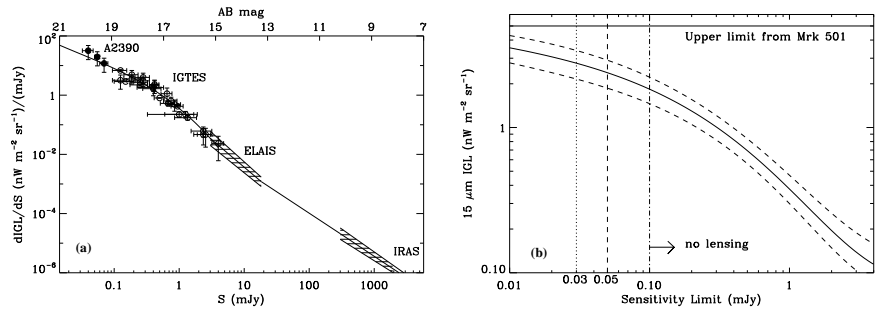}
\caption{{\bf a)} Differential contribution to the 15\,$\mu$m
Integrated Galaxy Light as a function of flux density and AB
magnitude. The plain line is a fit to the data: Abell 2390 (Altieri
\etal 1999), the ISOCAM Guaranteed Time Extragalactic Surveys (IGTES,
Elbaz \etal 1999), the European Large Area Infrared Survey (ELAIS,
Serjeant \etal 2000) and the IRAS all sky survey (Rush, Malkan \&
Spinoglio 1993). {\bf b)} Contribution of ISOCAM galaxies to the
15\,$\mu$m extragalactic background light (EBL), i.e. 15\,$\mu$m
Integrated Galaxy Light (IGL), as a function of sensitivity or $AB$
magnitude $\left(AB=\ -2.5\,log\left(S_{\rm
mJy}\right)+16.4\right)$. The plain line is the integral of the fit to
dIGL/dS (Fig.a).  The dashed lines correspond to 1-$\sigma$ error bars
obtained by fitting the 1-$\sigma$ upper and lower limits of dIGL/dS.}
\label{FIG:dEBL}
\end{figure}

The source counts at 15\mum exhibit a strong excess of faint sources
below $S_{15}\sim$ 2 mJy. This excess is usually defined by comparison
with model predictions assuming that galaxies behaved similarly in the
distant universe as they do today. Such ``no evolution'' behavior is
represented by a shaded area in the Fig.~\ref{FIG:dNdS} (see figure
caption). Galaxies above this flux density do fall within this region,
as illustrated by the data points from the ELAIS-S1 field (Gruppioni
\etal 2003, see also Oliver \etal in this book). In
Fig.~\ref{FIG:dNdS}, we have separated the data points from the IGTES
(Elbaz \etal 1999) between those below and above $S_{15}=$ 1 mJy, with
filled and open dots respectively.  Data above this flux density from
Elbaz \etal (1999) appear to be inconsistent with those derived from
ELAIS-S1. Most of those points were derived from the Shallow Survey of
the Lockman Hole within the IGTES, which suffered from having less
redundant observations of a given sky pixel. At that time, ISOCAM data
reduction methods were not optimized for such surveys, but since then,
they have been improved to better deal with such shallow surveys.
Among the techniques that are discussed below, the ``Lari'' technique
is particularly suitable for such low redundancy surveys (and even
more for the very shallow ELAIS surveys, see Oliver \etal in this
book) and a recent analysis of the Lockman Hole Deep (Fadda \etal
2004, large open squares) and Shallow (Rodighiero \etal 2004, large
open stars) surveys from the IGTES provided new number counts at these
flux densities perfectly consistent with those derived from ELAIS-S1
by Gruppioni \etal (2003) also using the same ``Lari'' technique. Note
that the models designed to fit the ISOCAM number counts were
constrained by the Elbaz \etal (1999) results, hence overproduce the
number of sources above $S_{15}\sim$ 2 mJy.  As a natural result, they
have also overpredicted the number of sources detected in the high
flux density regime at 24\mum with Spitzer (see Sect.~\ref{future} and
Papovich \etal 2004, Chary \etal 2004).

Above the Earth's atmosphere, the 15\,$\mu$m light is strongly
dominated by the zodiacal emission from interplanetary dust and it has
not yet been possible to make a direct measurement of the 15\,$\mu$m
background, or EBL. Individual galaxies contribute to this background
and a lower limit to the 15\,$\mu$m EBL can be obtained by adding up
the fluxes of all ISOCAM galaxies detected per unit area down to a
given flux limit. The resulting value is called the 15\,$\mu$m
integrated galaxy light (IGL).

As in Elbaz \etal (2002), the differential number counts can be
converted into a differential contribution to the 15\,$\mu$m IGL as a
function of flux density, estimated from the following formula:
\begin{equation}
\frac{dIGL}{dS}=~\frac{dN}{dS}\times~\left(\frac{S_{15}}{10^{20}}\right)~\times~\nu_{15}
\end{equation}
where $dN$(sr$^{-1}$) is the surface density of sources with a flux
density $S_{\nu}$[15\,$\mu$m]$=S_{15}$ (mJy) over a flux density bin
$dS$(mJy) (1 mJy= 10$^{-20}$ nW m$^{-2}$ Hz$^{-1}$) and $\nu_{15}$(Hz)
is the frequency of the 15\,$\mu$m photons.

About 600 galaxies below $S_{15}\sim$ 3 mJy, were used to provide the
points with errors bars in Fig.~\ref{FIG:dEBL}a. Fig.~\ref{FIG:dEBL}b
shows the 15\,$\mu$m IGL as a function of depth. It corresponds to the
integral of Fig.~\ref{FIG:dEBL}a, where the data below 3 mJy were
fitted with a polynomial of degree 3 and the 1-$\sigma$ error bars on
$dIGL/dS$ were obtained from the polynomial fit to the upper and lower
limits of the data points. The 15\,$\mu$m IGL does not converge above
a sensitivity limit of $S_{15}\sim$ 50\,$\mu$Jy, but the flattening of
the curve below $S_{15}\sim 0.4$ mJy suggests that most of the
15\,$\mu$m EBL should arise from the galaxies already unveiled by
ISOCAM. Above this flux density limit, where the completeness limit is
larger than 80\,\% - including lensed objects-, it is equal to 2.4
$\pm$ 0.5 nW m$^{-2}$ sr$^{-1}$ (Elbaz \etal 2002). Down to a 50\,\%
completeness limit, Metcalfe \etal (2003) found a 10\,\% larger value
including sources down to 30\,$\mu$Jy but the statistics remains
limited at these depths with only four sources below 50\,$\mu$Jy.

Franceschini et al. (2001) and Chary \& Elbaz (2001), developped
models which reproduce the number counts from ISOCAM at 15\,$\mu$m,
from ISOPHOT at 90 and 170\,$\mu$m and from SCUBA at 850\,$\mu$m, as
well as the shape of the CIRB from 100 to 1000\,$\mu$m. These models
consistently predict a 15\,$\mu$m EBL of:
\begin{equation}
EBL^{models}(15\,\mu {\rm m}) \sim 3.3~{\rm nW~m^{-2}~sr^{-1}}
\end{equation}

If this prediction from the models is correct then about $73\pm15\,\%$
of the 15\,$\mu$m EBL is resolved into individual galaxies by the
ISOCAM surveys. 

This result is consistent with the upper limit on the 15\,$\mu$m EBL
estimated by Stanev \& Franceschini (1998) of:
\begin{equation}
EBL^{max}(15\,\mu {\rm m}) \sim 5~{\rm nW~m^{-2}~sr^{-1}}
\end{equation}
This upper limit was computed from the 1997 $\gamma$-ray outburst of
the blazar Mkn 501 ($z=$ 0.034) as a result of the opacity of mid IR
photons to $\gamma$-ray photons, which annihilate with them through
$e^+e^-$ pair production. It was since confirmed by Renault et
al. (2001), who found an upper limit of 4.7 nW m$^{-2}$ sr$^{-1}$ from
5 to 15\,$\mu$m.

Nearly all the IGL is produced by sources fainter than 3 mJy
(94\,$\%$) and about 70\,$\%$ by sources fainter than 0.5 mJy. This
means that the nature and redshift distribution of the galaxies
producing the bulk of the 15\,$\mu$m IGL can be determined by studying
these faint galaxies only.

\subsection{Mid IR as a SFR indicator}
\label{MIRtracer}
\begin{figure}
\includegraphics[width=12cm]{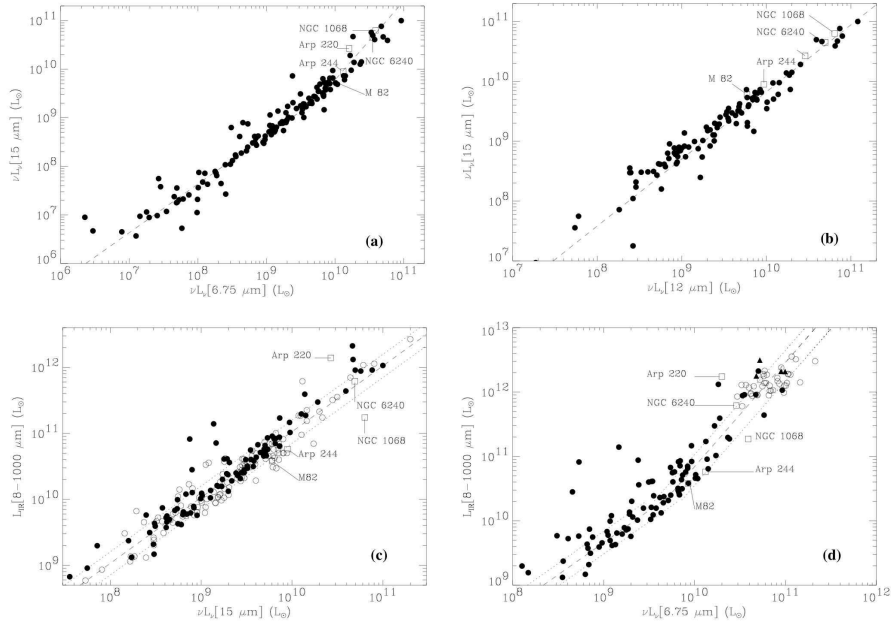}
\caption{IR luminosity correlations for local galaxies (from Elbaz
\etal 2002). {\bf a)} ISOCAM-LW3 (15\,$\mu$m) versus ISOCAM-LW2
(6.75\,$\mu$m) luminosities ($\nu L_{\nu}$) (56 galaxies). {\bf b)}
ISOCAM-LW3 (15\,$\mu$m) versus IRAS-12\,$\mu$m luminosities (45
galaxies). {\bf c)} $L_{\rm IR}$[8-1000\,$\mu$m] versus ISOCAM-LW3
(15\,$\mu$m) luminosity (120 galaxies). {\bf d)} $L_{\rm
IR}$[8-1000\,$\mu$m] versus LW2-6.75\,$\mu$m luminosities (91
galaxies). Filled dots: galaxies from the ISOCAM guaranteed time (47
galaxies including the open squares). Open dots: 40 galaxies from
Rigopoulou \etal (1999). Empty triangles: 4 galaxies from Tran et
al. (2001). Galaxies below L$_{\rm IR}\sim$ 10$^{10}$ L\sol present a
flatter slope and have $L_{\rm IR}$/$L_{B}<$ 1.}
\label{FIG:correl}
\end{figure}

\begin{figure} 
\includegraphics[width=9cm]{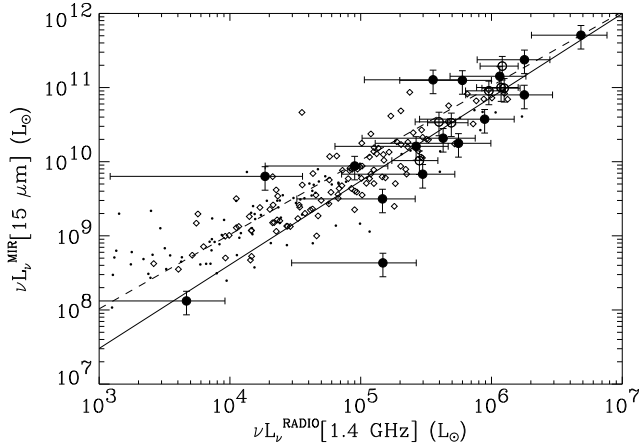}
\caption{15\,$\mu$m versus radio continuum (1.4 GHz) rest-frame
luminosities. Small filled dots: sample of 109 local galaxies from
ISOCAM and NVSS. Filled dots with error bars: 17 HDFN galaxies
($z\sim$0.7, radio from VLA or WSRT). Open dots with error bars: 7
CFRS-14 galaxies ($z\sim$0.7, Flores \etal 1999, radio from VLA). Open
diamonds: 137 ELAIS galaxies ($z\sim$ 0-0.4).}
\label{FIG:radio}
\end{figure}

When normalized to the 7.7\mum PAH (polycyclic aromatic hydrocarbon)
broad emission line, the spectra of different galaxies exhibit very
different 15, 25, 60 or 100\mum over 7.7\mum ratio. This was often
interpreted as an indication that measuring the monochromatic
luminosity of a galaxy at one mid IR wavelength was useless to
determine its bolometric IR luminosity, L\ir= L(8--1000\mum).
However, the variation of the far over mid IR ratio is correlated with
L\ir and local galaxies do exhibit a strong correlation of their mid
and far IR luminosities (Fig.~\ref{FIG:correl}). These correlations
can be used to construct a family of template SEDs or correlations
from which the L\ir, and therefore SFR, of a galaxy can be derived
from its mid IR luminosity (Chary \& Elbaz 2001, Elbaz \etal 2002).
The L\ir derived from this technique are consistent with those derived
by the radio-far IR correlation, when radio-mid-far IR data exist
(Elbaz \etal 2002, Garrett 2002, Gruppioni \etal 2003). In the
Fig.~\ref{FIG:radio}, we have reproduced the plot from Elbaz \etal
(2002) complemented with galaxies detected within the ELAIS survey
(Rowan-Robinson \etal 2004). Except at low luminosities were the
contribution of cirrus to the IR luminosity becomes non negligible,
the 1.4 GHz and 15\,$\mu$m rest-frame luminosities are correlated up
to $z\sim$ 1 and therefore predict very consistent total IR
luminosities. A similar result was later on obtained using the MIPS
instrument onboard the Spitzer Space Observatory at 24\mum (Appleton
\etal 2004).

Several studies compared the SFR derived from the IR luminosity with
the optical SFR derived from the H$\alpha$ emission line (Rigopoulou
\etal 2000, Cardiel \etal 2003, Flores \etal 2004, Liang \etal
2004). Rigopoulou \etal (2000) found a large excess of SFR(IR) versus
SFR(H$\alpha$) even after correcting the latter for extinction. The
Balmer decrement was only measured for limited number of objects in
the sample and the extinction correction was derived from broadband
photometry, which suffers from strong uncertainties in particular due
to the degeneracy between age, metallicity and extinction.  However,
Cardiel \etal (2003) confirmed the direct measure of the SFR(IR)
excess using the combination of NIRSPEC and LRIS, for the distant
galaxies, and the Echelle Spectrograph and Imager (ESI), for the
closer ones, at the Keck telescope. Using high resolution VLT-FORS2
spectra, Flores \etal (2004) and Liang \etal (2004) were able to
measure directly the Balmer decrement (using H$\alpha$/H$\beta$ or
H$\beta$/H$\gamma$) and to subtract the underlying nebular emission
lines with a fit of the stellar continuum. Although the SFR(IR)
exceeds the estimate from emission lines for the most active objects,
the data present a clear correlation between SFR(IR) and
SFR(H$\alpha$) which suggests that the star formation regions
responsible for the IR luminosity of distant LIRGs are not completely
obscured. This correlations also confirms that the mid IR is indeed a
good SFR estimator.

\subsection{Nature of the ISOCAM Galaxies}
\label{natureCAM}
\begin{figure*}
\includegraphics[width=12cm]{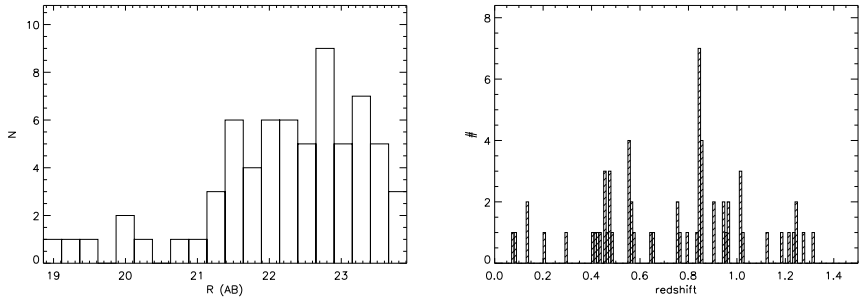}
\caption{(left) Histogram of the R(AB) magnitudes of the 15\mum ISOCAM
galaxies detected in the HDFN$+$FF (revised catalog of Aussel \etal
down to $\sim$ 30 $\mu$Jy. (right) Redshift distribution of the
HDFN$+$FF ISOCAM galaxies.}
\label{FIG:zdist}
\end{figure*}

\begin{figure*}
\includegraphics[width=12cm]{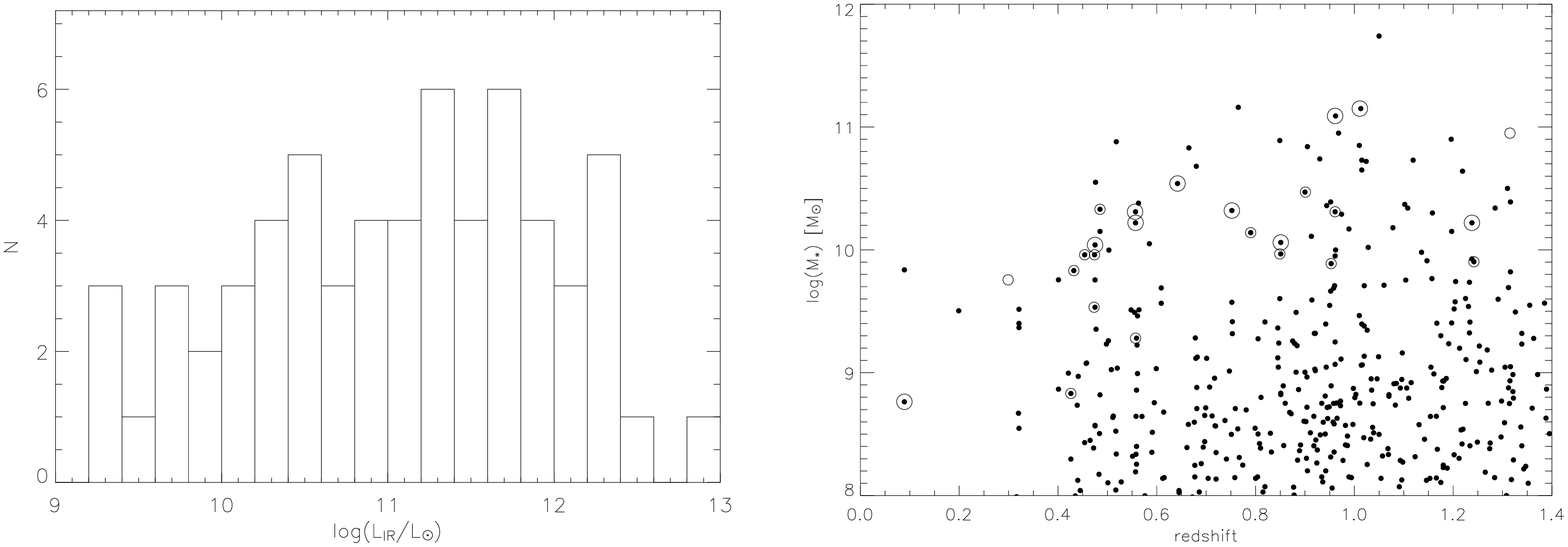}
\caption{(left) Distribution of the log(L\ir(8-1000\,\mum)/L\sol) of
the HDFN$+$FF ISOCAM galaxies derived from ther 15\mum flux
densities. (right) Stellar mass as a function of redshift of field
HDFN proper galaxies (dark dots) and of ISOCAM galaxies (open
circles).  All stellar masses were derived using multi-bands SED
modelling by Dickinson \etal (2003).}
\label{FIG:mass}
\end{figure*}

The Hubble Deep Field North and its Flanking Fields (HDFN$+$FF)
provides the best coverage in spectroscopic redshifts and deep optical
images for an ISOCAM deep survey. We used the revised version of the
Aussel \etal (1999) catalog for which 85\,\% (71\,\%) of the 40 (86)
galaxies above 100\,$\mu$Jy (30\,$\mu$Jy) have a spectroscopic
redshift to determine the average properties of ISOCAM galaxies
summarized in the Figs.~\ref{FIG:zdist},~\ref{FIG:mass}. Their optical
counterparts are relatively bright and their median-mean redshift is
close to $z\sim$0.8 (Fig.~\ref{FIG:zdist}). Note the redshift peaks in
which the ISOCAM galaxies are located, leaving wide empty spaces in
between. Most ISOCAM galaxies are located within large-scale
structures, here mainly those at z=0.848 and z=1.017, which might be
galaxy clusters in formation where galaxy-galaxy interactions are
amplified (Elbaz \& Cesarsky 2003, see discussion below).

Thanks to the deepest soft to hard X-ray survey ever performed with
Chandra in the HDFN, it is possible to pinpoint active galactic nuclei
(AGNs) in this field including those affected by dust extinction. Only
five sources were classified as AGN dominated on the basis of their
X-ray properties (Fadda et al. 2002). Hence, unless a large number of
AGNs are so dust obscured that they were even missed with the 2
Megaseconds Chandra survey, the vast majority of ISOCAM galaxies are
powered by star formation. This result is consistent with observations
of local galaxies which indicate that only the upper luminosity range
of ULIRGs are dominated by an AGN (Tran \etal 2001).

Using the mid-far IR correlations (Chary \& Elbaz 2001, Elbaz \etal
2002, see also Sec.~\ref{MIRtracer}), the L\ir distribution of the
HDFN mid IR sources is plotted in Fig.~\ref{FIG:mass}a. Most of them
belong to the LIRG and ULIRG regime, although when including flux
densities below completeness down to 30\,$\mu$Jy, one finds also
intermediate luminosities. Finally, their stellar masses are among the
largest in their redshift range, when compared to the stellar mass
estimates by Dickinson \etal (2003) in the HDFN
(Fig.~\ref{FIG:mass}b).

\begin{figure} 
\includegraphics[width=12cm]{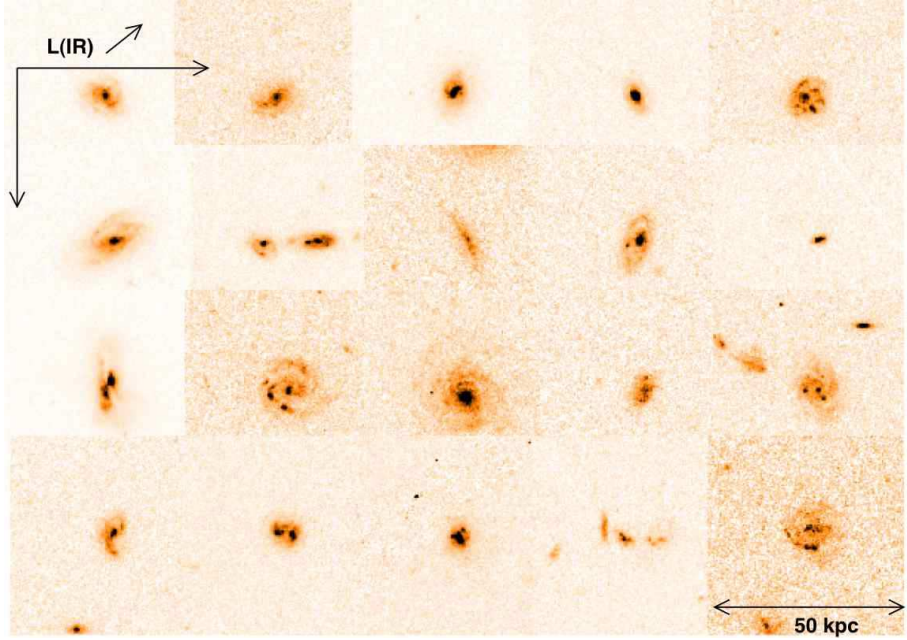}
\caption{HST-ACS images of LIR galaxies with 11 $\leq$ log(L$_{\rm
IR}/L_{\odot}$) $\leq$ 12 (LIRGs) and $z\sim$ 0.7. The double-headed
arrow indicates the physical size of 50 kpc. The IR luminosity
increases from left to right and from top to bottom.}
\label{FIG:camgal}
\end{figure}

In the local universe, both LIRGs and ULIRGs exhibit the typical
morphology of major mergers, i.e. mergers of approximately equal mass
(Sanders \& Mirabel 1996, Sanders, Surace \& Ishida 1999). In the case
of LIRGs, the merging galaxies show a larger separation than for
ULIRGs, which are mostly in the late phase of the
merger. Fig.~\ref{FIG:camgal} (Elbaz \& Moy 2004) presents the HST-ACS
morphology of a sample of $z\sim$ 0.7 LIRGs in the GOODSN field
(extended HDFN). Less than half of these galaxies clone the morphology
of local LIRGs, which implies that the physical processes switching on
the star formation activity in distant LIRGs might be different than
for local ones. The gas mass fraction of younger galaxies being
larger, other types of interactions might generate a LIRG phase in the
distant universe, such as minor mergers or even passing-by galaxies
producing a tidal effect. The fact that such interactions are more
frequent than major mergers could also explain the importance of the
LIRG phase for galaxies in general and also the possibility for a
galaxy to experience several intense bursts in its lifetime. The
appearance of this phase of violent star formation could be
facilitated during the formation of groups or clusters of galaxies.  

A striking example of this is given by the large fraction of LIRGs
detected in the distant galaxy cluster J1888.16CL, located at a
redshit of $z=$ 0.56 (Duc \etal 2004). Among the 27 objects for which
spectra were obtained, six of them belong to the cluster while an
extra pair with slightly higher redshifts may lie in infalling groups.
All eight galaxies exhibit weak emission lines in their optical
spectra, typical of dust enshrouded star forming galaxies, none of
these lines being broad enough to indicate the presence of type I
AGNs.  In this relatively young galaxy cluster, the mechanism that may
be triggering SFRs between 20 and 120 M$\odot$ yr$^{-1}$ for at least
eight objects of the cluster could well be tidal collisions within
sub-structures or infalling groups. On a more local scale, the mid IR
emission of the Abell 1689 ($z=$ 0.181) galaxies exhibits an excess of
the B-[15] color with respect to richer and closer galaxy clusters,
such as Coma and Virgo, which suggests the presence of a mid IR
equivalent to the Butcher-Oemler effect, i.e. the star formation
activity of galaxies as reflected by their mid IR emission increases
with increasing redshift (Fadda \etal 2000). The high fraction of blue
galaxies intially reported for this cluster by Butcher \& Oemler
(1984) was later on confirmed by Duc \etal (2002), who also found that
the actual SFR for these galaxies was on average ten times larger than
the one derived from the [OII] emission line implying that 90\,\% of
the star formation activity taking place in this cluster was hidden by
dust.

\begin{figure} 
\includegraphics[width=12cm]{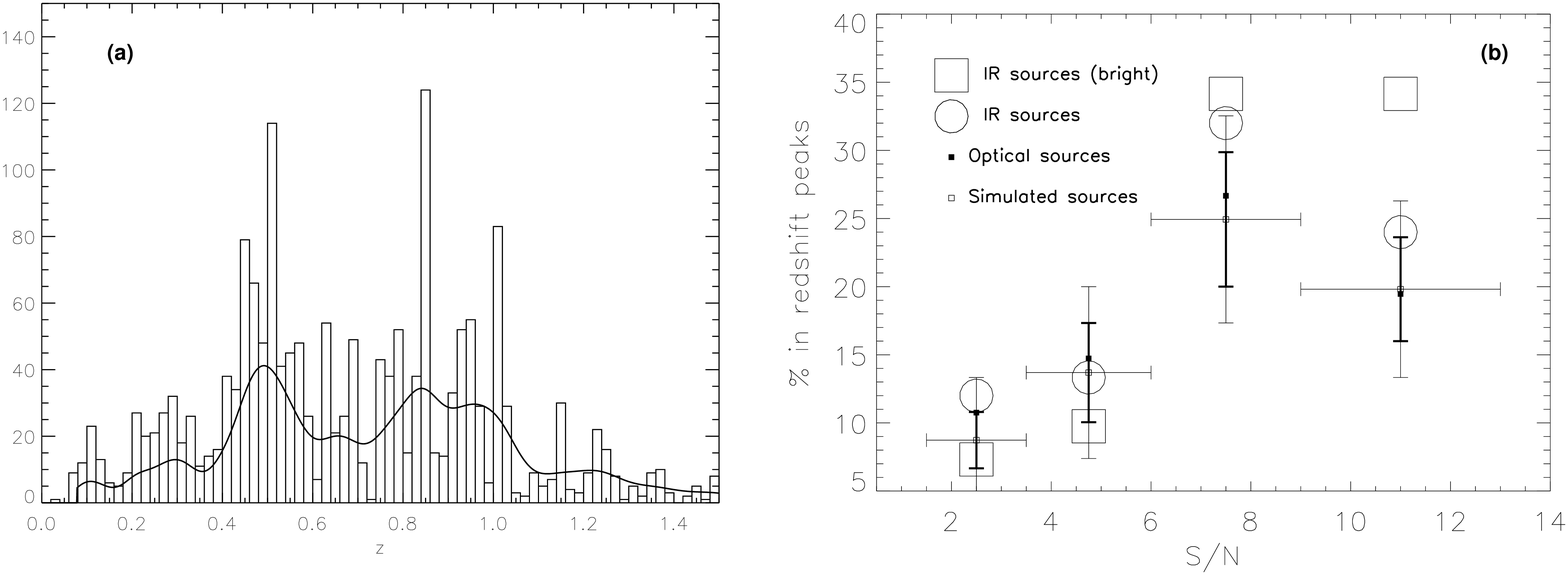}
\caption{{\bf (a)} Redshift distribution of field galaxies in the
GOODSN field. The continuous line is the smoothed distribution with a
window of 15,000 km/s. {\bf (b)} Differential fraction of sources
within redshift peaks (see definition in the text) stronger than
N-$\sigma$. Small filled squares: fraction of sources in peaks for the
whole optical catalog of 930 field galaxies (Wirth \etal 2004). Small open
squares: median of the fraction of sources in redshift peaks for a
series of Monte Carlo simulations of a sub-sample of the field
galaxies corresponding to the same number of galaxies as in the ISOCAM
catalog and within the same range of redshifts and optical-near IR
magnitudes. Error bars contain 68 and 90\,\% of the simulations. Large
Open Circles: total sample of ISOCAM galaxies (75 sources with
spectroscopic redshifts and optical-near IR magnitudes). Large open
squares: sub-sample of ISOCAM galaxies above the completeness limit of
0.1 mJy (41 sources).}
\label{FIG:MCpeaks}
\end{figure}

LIRGs could then be a tracer of large-scale structures in formation as
suggested by their redshift distribution (Fig.~\ref{FIG:zdist}b, see
also Elbaz \& Cesarsky 2003). In order to test this hypothesis, Moy \&
Elbaz (in prep.) compared the fraction of ISOCAM galaxies (above the
completeness limit of 0.1 mJy) found in ``redshift peaks'' to the one
obtained when randomly selecting field sources of equal optical and
K-band magnitudes in the same redshift range. Redshift peaks of
different strengths, measured as N-$\sigma$, were defined by smoothing
the field galaxies redshift distribution by 15,000 km/s
(Fig.\ref{FIG:MCpeaks}a) and measuring the peaks N-$\sigma$ above the
smoothed distribution.  The Monte-Carlo samples of field galaxies to
be compared to the redshift distribution of the ISOCAM galaxies were
selected within the real redshift distribution and not the smoothed
one. The distribution of the Monte-Carlo simulations are shown with
error bars at the 68 and 90\,\% level. When considering the whole
ISOCAM catalog, i.e. including faint sources below completeness, it
appears that ISOCAM galaxies are more clustered than field galaxies.
Less than 32\,\% of the simulations present a comparable clustering
than the whole ISOCAM sample which contains a large fraction of non
LIRGs.  Strickingly, when only selecting the brightest galaxies above
the completeness limit of 0.1 mJy, one finds that they are LIRGs and
ULIRGs which fall in the densest redshift peaks, above 6-$\sigma$. The
probability to randomly select a sample of galaxies from the field
(with equal magnitudes and redshift range) in the redshift peaks of
6-$\sigma$ and above is less than 1\,\%. As a result, mid IR surveys
are very efficient in selecting over-dense regions in the universe,
which in return are very efficient in producing a LIRG. On the
contrary, mid IR selected galaxies are locally less clustered than
field galaxies (Gonzalez-Solares \etal 2004), which could be a natural
result of the fact that only the less clustered galaxies still produce
IR luminous phases while more clustered galaxies lived their IR
luminous phase in the past (see also Elbaz \& Cesarsky 2003).

Finally, one question remains to be addressed about distant LIRGs: how
long does this starburst phase last and how much stellar mass is
produced during that time ? Marcillac \etal (2004) used a bayesian
approach and simulated 200,000 virtual high-resolution spectra with
the Bruzual \& Charlot (2003) code to determine the recent star
formation history of distant LIRGs as well as their stellar
masses. These ISOCAM galaxies were observed using the VLT-FORS2
($\Delta \lambda$/$\lambda \sim$ 2000 in the rest-frame) in three
different fields. A prototypical LIRG at $z\sim$0.7 is found to have a
stellar mass of $\sim$5$\times$10$^{10}$ M\sol and to produce about
10\,\% of this stellar mass within about 10$^8$ years during the
burst. A remarkable result of this study is that the position of
distant LIRGs in a diagram showing the value of the H8 Balmer
absorption line equivalent width versus the strength of the
4000\,\AA\, break signs the presence of a burst of star formation
within these galaxies, with an intensity of about 50 M\sol yr$^{-1}$
as also derived from their mid IR emission. This result supports the
idea that distant LIRGs are not completely opaque to optical light and
that one can learn something about their star formation history based
on their optical spectra.

Liang \etal (2004) compared the gas metallicity of the same sample of
objects than Marcillac \etal (2004) with local galaxies of similar
absolute magnitudes in the B band. Even accounting for an evolution in
the B-band luminosity, the distant LIRGs turn out to be about twice
less metal rich. This result suggests that between $z\sim$1 and today,
LIRGs do produce a large fraction of the metals located in their host
galaxies in agreement with the strong evolution of the cosmic star
formation history found by the models fitting the ISO source counts.

\section{Cosmic evolution, star formation rate history}
\label{csfr}
The combination of surveys at different wavelengths, from ISOCAM,
ISOPHOT and SCUBA, together with the the shape and intensity of the
CIRB, was used by several authors to constrain the parameters of their
backward evolution models assuming a combination of luminosity and
density evolution as a function of redshift of the IR luminosity
function at 15 or 60\mum: 
Roche \& Eales (1999),
Tan \etal (1999),
Devriendt \& Guiderdoni (2000),
Dole \etal (2000),
Chary \& Elbaz (2001),
Franceschini \etal (2001, 2003),
Malkan \& Stecker (2001),
Pearson (2001),
Rowan-Robinson (2001), King \& Rowan-Robinson (2003),
Takeuchi \etal (2001),
Xu \etal (2001,2003),
Balland \etal (2002),
Lagache, Dole \& Puget (2003),
Totani \& Takeuchi (2002),
Wang (2002).

\begin{figure}
\includegraphics[width=9cm]{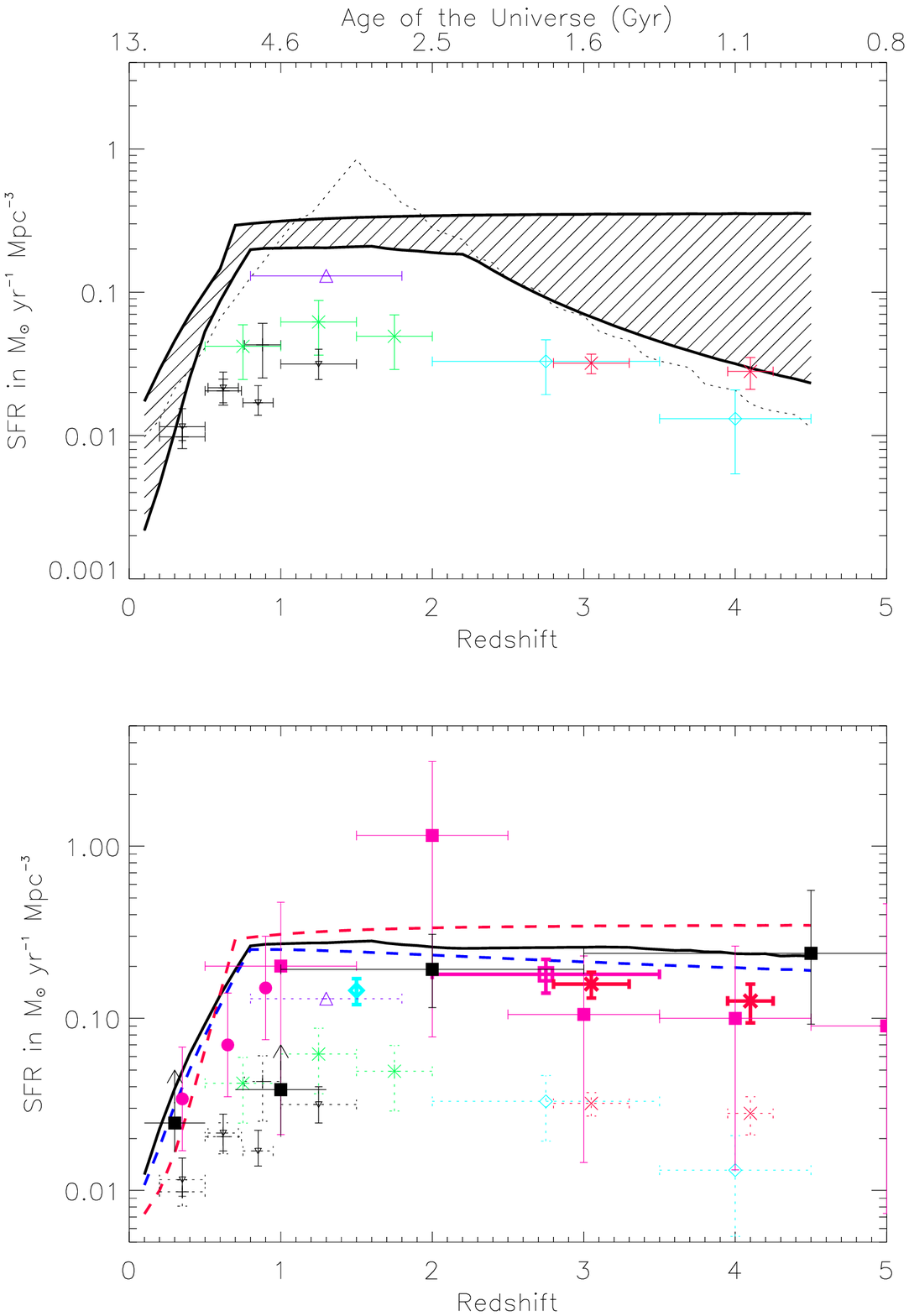}
\caption{Star Formation Rate from Chary \& Elbaz (2001). Upper panel:
min and max range from their model, and observed UV/opt data; dots
represent Xu \etal (2001) model.
Lower panel: 3 different evolution scenarios from their model and data
corrected for extinction. Line: pure luminosity; upper dash: pure
density; lower dash: luminosity+density.}
\label{FIG:csfr_chary}
\end{figure}

Being limited by the sensitivity of the extragalactic surveys, the
major output of these models was to show that LIRGs and ULIRGs were
much more common in the past than they are today. Chary \& Elbaz
(2001) derived that the comoving IR luminosity due to LIRGs was about
seventy times larger at $z\sim$1 than it is today
(Fig.~\ref{FIG:csfr_chary}). Quite logically, it resulted that the
contribution of LIRGs to the cosmic star formation history was so
large in the past that it dominated the integrated star formation that
galaxies experienced in the past. Hence, LIRGs should now be
considered not as a type of galaxies but instead as a common phase of
intense star formation that any galaxy may have experienced.

\begin{figure}
\includegraphics[width=9cm]{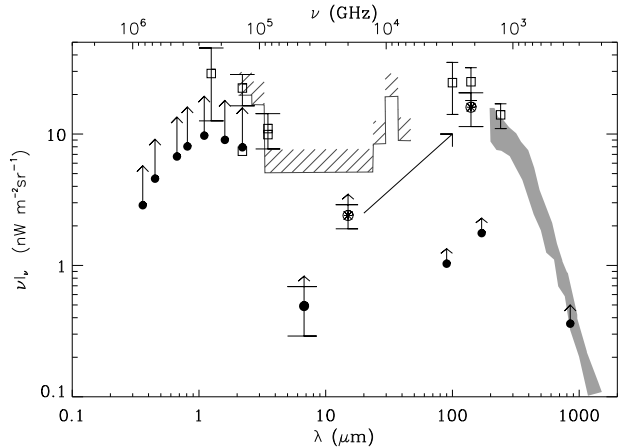}
\caption{Integrated Galaxy Light (IGL, filled dots) and Extragalactic
Background Light (EBL, open squares, grey area) from the UV to
sub-millimeter (from Elbaz \etal 2002). EBL measurements from COBE:
200-1500\,$\mu$m EBL from COBE-FIRAS (grey area, Lagache \etal 1999),
1.25, 2.2, 3.5, 100, 140\,$\mu$m EBL from COBE-DIRBE (open
squares). IGL in the U,B,V,I,J,H,K bands from Madau \& Pozzetti
(2000). The upper end of the arrows indicate the revised values
suggested by Bernstein et al. (2001, factor two higher). 6.75\,$\mu$m
(ISOCAM-LW2 filter) IGL from Metcalfe et al. (2003, filled dot with
error bar and arrow). Hatched upper limit from Mkn 501 (Stanev \&
Franceschini 1998). The ISOCAM 15\,$\mu$m IGL (2.4 $\pm$ 0.5 nW
m$^{-2}$ sr$^{-1}$) is marked with a star surrounded by a circle. The
other star surrounded by a circle is the prediction of the
contribution of the 15\mum sources to 140\mum (Elbaz \etal 2002).}
\label{FIG:cirb}
\end{figure}

Elbaz \etal (2002) derived the contribution to the peak of the CIRB at
140\mum from the population of LIRGs around $z\sim$ 0.8 detected in
the ISOCAM 15\mum deep surveys and found that these objects alone
can explain more than two thirds of the peak and integrated intensities
of the CIRB (see Fig.~\ref{FIG:cirb}). Hence the CIRB is the signature
of the strong redshift evolution of LIRGs and the fossil record
of star formation which took place in such burst phases.

\section{From ISO to Spitzer, Herschel, the JWST and ALMA}
\label{future}
\begin{figure}
\includegraphics[width=9cm]{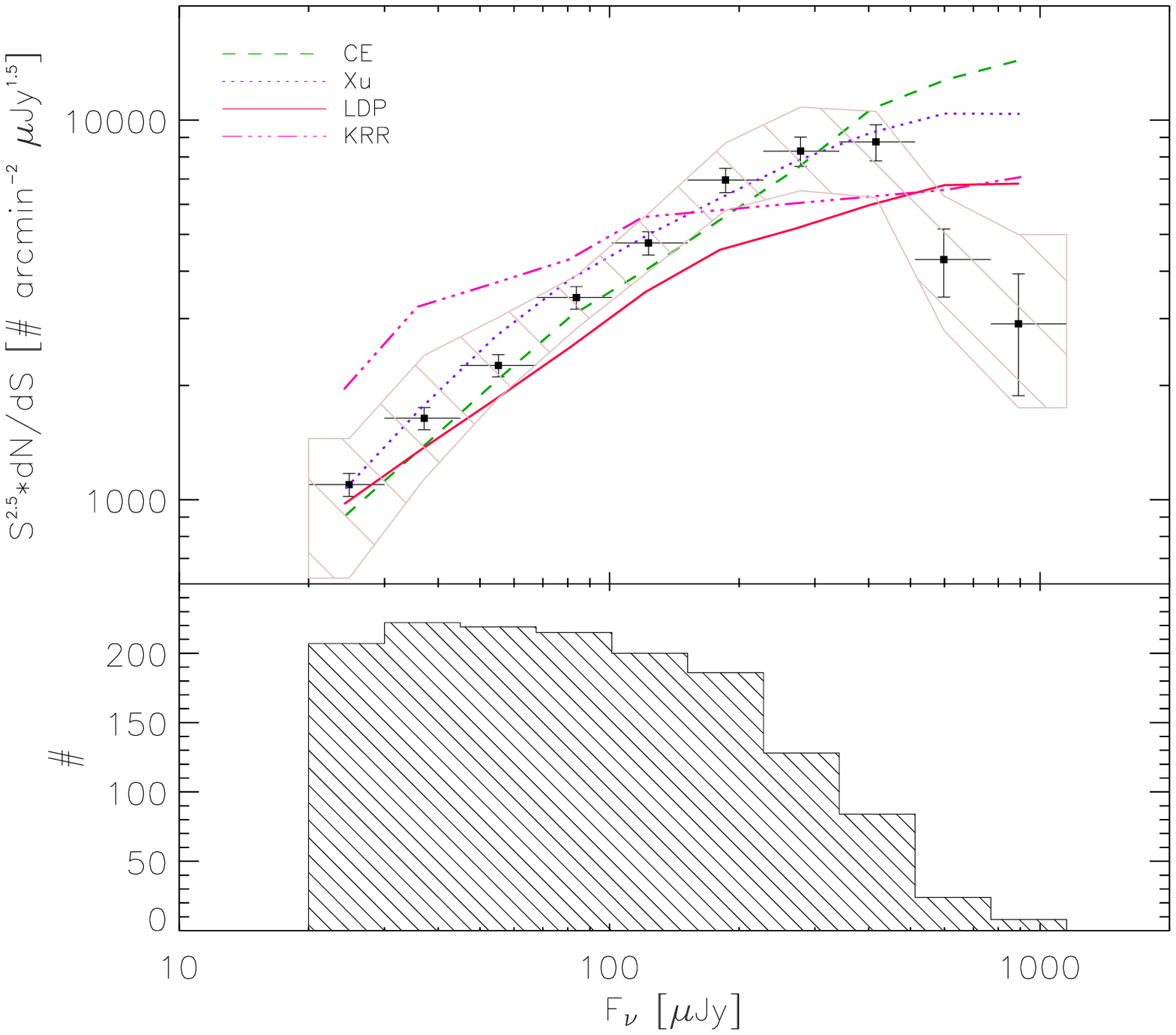}
\caption{Completeness corrected galaxy counts in the MIPS 24\mum
channel from Spitzer observations of the ELAIS-N1 field (from Chary
\etal 2004). The error bars reflect the Poissonian uncertainty.  The
horizontal bars represent the minimum and maximum flux density in that
bin.  The lines show four models for 24\mum counts: King \&
Rowan-Robinson (2003, KRR), Xu \etal (2001, Xu), Chary \& Elbaz (2001,
CE), Lagache, Dole \& Puget (2003, LDP). The symbols are plotted at
the average of the flux densities of the detected sources in that bin
for the data while the lines are plotted at the counts-weighted flux
average for the models. The lower plot in the figure shows the
histogram of the actual number of sources detected in each flux bin
without any completeness correction. }
\label{FIG:spitzer}
\end{figure}

\begin{figure}
\includegraphics[width=9cm]{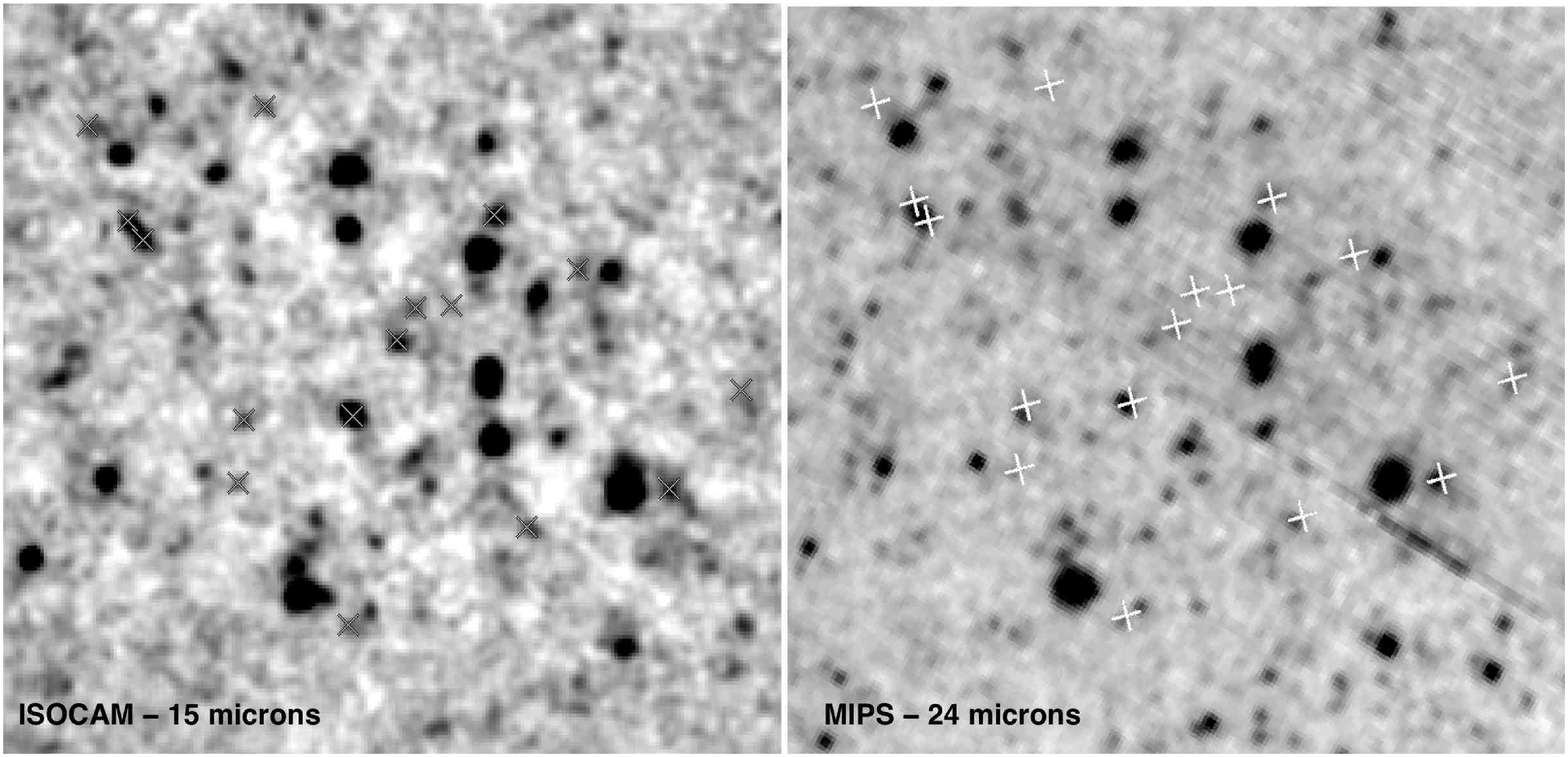}
\caption{ISOCAM 15\mum image (left) of the Ultra-Deep Survey in the
Marano FIRBACK field (depth 140\,$\mu$Jy, 80\,\% completeness) versus
Spitzer MIPS-24\mum image (right; depth 110\,$\mu$Jy, 80\,\%
completeness). The crosses identify 16 galaxies detected at 15 and
24\mum for which VLT-FORS2 spectra were obtained and which SEDs were
fitted in Elbaz \etal (2004).}
\label{FIG:image_1524}
\end{figure}

On August 23rd, 2003, NASA's Spitzer space telescope (formerly SIRTF)
was launched.  Among its first results came the source counts at
24\mum down to $\sim$ 20\,$\mu$Jy which confirmed what ISO deep
surveys already saw: a strong excess of faint sources indicating a
rapid redshift evolution of IR luminous galaxies. When compared to
models developped to fit the ISO counts, the faint end of the Spitzer
counts are perfectly fitted as shown in the Fig.~\ref{FIG:spitzer}
reproduced from Chary \etal (2004).  On the high flux density range,
around 1 mJy and above, less galaxies are found than predicted by
those models (see also Papovich \etal 2004). This is partly, if not
integrally, due to the fact that even ISOCAM-15\mum number counts were
initially overestimated above $S_{15}\sim$ 1 mJy as discussed in
Sect.~\ref{camcounts}, since the data reduction techniques were not
optimized for the surveys with little redundancy over a given sky
pixel. Although some refinement of the template SEDs used in the
models might be considered (as suggested by Lagache \etal 2004), the
24\mum number counts appear to be perfectly consistent with the
up-to-date 15\mum counts. Moreover, the high flux density range does
not strongly affect the conclusions of the models based on previous
15\mum counts since bright objects do not contribute significantly to
the CIRB and to the cosmic density of star formation. Hence these
refinements are not strongly affecting the conclusions derived on
galaxy formation and evolution based on the ISO deep surveys and
summarized in the previous sections (see also Dole \etal 2004 for the
Spitzer counts in the far IR).

Another hint on the consistency of Spitzer MIPS-24\mum surveys with
ISOCAM-15\mum is given by the comparison of the images themselves.
Galaxies detected at 15 and 24\mum are clearly visible in both images
in Fig.~\ref{FIG:image_1524}, although several 24\mum sources do not
have a 15\mum counterpart. This results from the combination of the
better sensitivity of MIPS, by a factor 2 or slightly more for the
deepest surveys (galaxies are detected down to $S_{15}\sim$ 40 $\mu$Jy
in the HDFN, see Aussel \etal 1999, without lensing magnification),
and of the k-correction. For galaxies above $z\sim$ 1, the PAH bump
centered on the PAH feature at 7.7\mum starts to exit to 15\mum
broadband while it remains within the MIPS-24\mum band up to $z\sim$
2. The combination of ISOCAM and MIPS can be used to test whether the
library of template SEDs that were used to derive ``bolometric'' IR
luminosities from 8 to 1000\mum for the 15\mum galaxies are correct at
least in the mid IR range. One of the most important test is to check
whether the 7.7\mum PAH bump is still present at $z\sim$ 1 and whether
the 24 over 15\mum flux density ratio is consistent with the template
SEDs used by the models from which star formation histories were
derived. The template SEDs designed by Chary \& Elbaz (2001) provide a
very good fit to the combination of both mid IR values for a sample of
16 galaxies detected by ISOCAM and MIPS (crosses on the
Fig.~\ref{FIG:image_1524}). The $L_{\rm IR}$(8-1000\mum) derived from
either the 15 or 24\mum luminosities or the combination of both to
constrain the SED fit present a 1-$\sigma$ dispersion of only 20\,\%
(Elbaz \etal 2004). For galaxies located around $z\sim$ 1, the
relative 15 and 24\mum luminosities clearly suggest the presence of a
bump at 7.7\mum as observed in nearby galaxies and due to PAHs.

Many questions remain unsolved that will be addressed by future
missions, staring with Spitzer. Only when a fair sample of redshifts
will have been determined for the distant LIRGs will we be able to
definitely ascertain the redshift evolution of the IR luminosity
function.  Already for the brightest part of it, campaigns of redshift
measurements have started, on ELAIS fields for the nearby objects, and
on the Marano field with VIMOS and the Lockman Hole with DEIMOS for
more distant objects.  The fields selected for Spitzer legacy and
garanteed time surveys were also carefully selected to be covered at
all wavelengths and followed spectroscopically, so that this issue
should be addressed in the very near future. Due to confusion and
sensitivity limits, direct observations in the far IR will not reach
the same depth than mid IR ones until the launch of Herschel scheduled
for 2007. The direct access to the far IR distant universe with
Herschel will certainly bring major information on galaxy formation,
together with the James Webb Space Telescope (JWST) up to 30\mum and
the Atacama Large Millimeter Array (ALMA), which will bring an
improved spatial resolution for the $z\sim2$ and above universe.
Among the questions to be solved, we do not resist to the temptation
of listing some of our favorite ones: what is the connection of
large-scale structure formation with LIRG phases in galaxies ? Can we
probe the formation of distant clusters by the detection of the epoch
when galaxies formed stars in such intense starbursts, producing
galactic winds and enriching the intra-cluster medium with metals ?
Have we really resolved the bulk of the hard X-ray background, which
peaks around 30 keV, and not left unknown some deeply buried AGNs
which could make a larger than 20\,\% fraction of the mid IR light
from distant LIRGs ?  What is the future of a distant LIRG, is it
producing stars in a future bulge or disk ? How uncertain is the
interpolation to lower luminosities than those observed done in the
models used to derive cosmic star formation history scenarios ?  This
is of course only an example of a vast series of questions which
indicate that the field opened for a large part by ISO has a long life
to come.

\end{article}
\end{document}